\documentclass[
reprint,
superscriptaddress,
amsmath,
amssymb,
aps,
nofootinbib
]{revtex4-2}

\usepackage{orcidlink}
\usepackage{amsmath}
\usepackage{hyperref}
\usepackage{mathtools}
\usepackage{graphicx}
\usepackage{color}
\usepackage{soul}
\sethlcolor{yellow}
\usepackage{tensor}
\usepackage{bm}

\def\sC{{\ensuremath{\mathcal{C}}}}
\def\sD{{\ensuremath{\mathcal{D}}}}

\def\sG{{\ensuremath{\mathcal{G}}}}

\def\sM{{\ensuremath{\mathcal{M}}}}

\def\sS{{\ensuremath{\mathcal{S}}}}
\def\sW{{\ensuremath{\mathcal{W}}}}

\usepackage{soul}

\begin{document}

\title{Formulating the complete initial boundary value problem in numerical relativity to model black hole echoes}

\author{Conner Dailey \orcidlink{0000-0003-2488-3461}
}
\email[Corresponding Author: ]{cdailey@pitp.ca}
\affiliation{Perimeter Institute for Theoretical Physics, Waterloo, ON, N2L Y25, Canada}
\affiliation{Waterloo Centre for Astrophysics, University of Waterloo, Waterloo, ON, N2L 3G1, Canada}
\affiliation{Department of Physics and Astronomy, University of Waterloo, Waterloo, ON, N2L 3G1, Canada}

\author{Erik Schnetter \orcidlink{0000-0002-4518-9017}}
\affiliation{Perimeter Institute for Theoretical Physics, Waterloo, ON, N2L Y25, Canada}
\affiliation{Department of Physics and Astronomy, University of Waterloo, Waterloo, ON, N2L 3G1, Canada}
\affiliation{Center for Computation \& Technology, Louisiana State University, Baton Rouge, Louisiana, USA}

\author{Niayesh Afshordi \orcidlink{0000-0002-9940-7040}}
\affiliation{Waterloo Centre for Astrophysics, University of Waterloo, Waterloo, ON, N2L 3G1, Canada}
\affiliation{Department of Physics and Astronomy, University of Waterloo, Waterloo, ON, N2L 3G1, Canada}
\affiliation{Perimeter Institute for Theoretical Physics, Waterloo, ON, N2L Y25, Canada}

\date{2024-09-10}

\begin{abstract}
In an attempt to simulate black hole echoes (generated by potential quantum-gravitational structure) in numerical relativity, we recently described how to implement a reflecting boundary outside of the horizon of a black hole in spherical symmetry. Here, we generalize this approach to spacetimes with no symmetries and implement it numerically using the generalized harmonic formulation. We cast the evolution equations and the numerical implementation into a Summation By Parts (SBP) scheme, which seats our method closer to a class of provably numerically stable systems. We implement an embedded boundary numerical framework that allows for arbitrarily shaped domains on a rectangular grid and even boundaries that evolve and move across the grid. As a demonstration of this framework, we study the evolution of gravitational wave scattering off a boundary either inside, or just outside,  the horizon of a black hole. This marks a big leap toward the goal of a generic framework to obtain gravitational waveforms for behaviors motivated by quantum gravity near the horizons of merging black holes.
\end{abstract}

\maketitle

\section{Introduction}

Recently we described a method for simulating black hole echoes in numerical relativity by imposing reflecting boundary conditions (BCs) near the horizon of a black hole in spherical symmetry \cite{Dailey_2023}. In that work, we defined reflecting BCs on a scalar field based on conservation laws and implemented it using modern numerical methods, including Summation By Parts (SBP) finite differencing operators and BCs applied using Simultaneous Approximation Terms (SATs), to ensure the system remains numerically stable. In an effort to apply this approach to spacetimes with no symmetries, we consider here a generalization to three spatial dimensions. 

The motivation to characterize and demonstrate stable initial boundary value problems (IBVPs) in numerical relativity comes from our interest in simulating black hole echoes \cite{Oshita:2019sat,Wang:2018gin,Ikeda:2021uvc,Wang:2019rcf,Abedi:2018npz,Abedi:2020sgg,Abedi:2016hgu,Burgess:2018pmm,Cardoso:2019apo}. 
Additionally, this is of direct interest to the numerical relativity community in the wake of the first demonstration of Cauchy characteristic matching (CCM) \cite{Ma_2024}, a technique that combines Cauchy evolution and characteristic evolution to provide gravitational wave (GW) scattering feedback to the sources in the Cauchy domain. In \cite{Ma_2024}, it was shown that it is possible to shrink the Cauchy domain while still obtaining accurate gravitational waveforms at null infinity by demonstrating several simulations of black holes perturbed by GWs. Shrinking the Cauchy domain can dramatically save on the computational cost of simulations, as most of this cost comes from the Cauchy evolution. Simulating black hole echoes and evolving CCM necessitate a numerically stable way of evolving dynamical spacetimes with boundaries in the strong gravity regime.

Abandoning spherical symmetry adds a considerable amount of complexity to the BCs and the manner in which they are applied. In \cite{Dailey_2023}, we derived and demonstrated reflecting BCs based on the Misner-Sharp mass, which happens to be locally conserved in spherical symmetry. However, in general, local conservation laws do not exist in general relativity. Instead, we have to rely on quasi-local conservation laws to define what it means to reflect gravitational radiation and matter waves. For a review of local and quasi-local conservation laws, see \cite{McGrath_2012}. Although we plan to develop BCs based on controlling the flux of a conserved quasi-local quantity, the focus in this work will be on the adjusted evolution equations, BC framework, and numerical implementation that we develop to make them feasible.

In general, the boundaries of the computational domain can have an arbitrary shape that may not always align with the coordinate directions chosen for the evolution. This is reminiscent of the problem of excision in numerical relativity, where a dynamical black hole interior must be excised so that a physical singularity does not enter the computational domain. Such an excision becomes straightforward with a numerical implementation that allows for an arbitrarily shaped boundary. If a formulation of numerical relativity with \emph{causal} characteristic speeds is used, no BCs are needed as long as the boundary remains within the apparent horizon of the black hole\footnote{Although being inside the apparent horizon is sufficient, it may not be strictly necessary since the apparent horizon world tube is still inside the event horizon.}.
In the case where the boundary is \emph{outside} of the apparent horizon, BCs are needed, and their numerical implementation becomes critical to the numerical stability of the simulation. We shall rely on modern methods to ensure that the simulation remains numerically stable for all time. These methods include:

\begin{itemize}
    \item SBP finite differencing operators for embedded boundaries
    \item SAT (or weakly implemented) BCs
    \item Symmetric hyperbolic formulations of numerical relativity, with an approximately conserved energy estimate
\end{itemize}

In \cite{Dailey_2023}, we used the Einstein-Christoffel formulation, as it is one of the few well-tested symmetric hyperbolic formulations of numerical relativity \cite{EC_Formulation}. However, in order to cast the evolution equations into an SBP scheme, we required a formulation more akin to the generalized harmonic (GH) formulation \cite{Garfinkle2002, Szil_gyi_2003, Friedrich1985,Lindblom_2006,Brown_2011}, which we consider here instead. Taking care that the energy estimate that comes from symmetric hyperbolicity is conserved in the sense of SBP can ensure that the energy estimate is bounded, and thus the simulation is, in principle, stable for all time (at least for linear short-wavelength perturbations). The principal goal of this work is to adapt the first order system introduced in \cite{Lindblom_2006} into a proper SBP-SAT scheme, which will require critical differences in the way the equations are evolved and the BCs are applied.

In Section \ref{Sec:CovSBP}, we define covariant SBP methods with the wave equation as an example. In Section \ref{Sec:GenHarmonic} we present our altered GH first order system and describe its SBP energy estimates. In Section \ref{Sec:ICs}, we describe how we set up initial conditions and in Section \ref{Sec:BCs} we present our BC framework. In Section \ref{Sec:EmbeddedMethods} we describe two embedded boundary SBP methods and chose one to implement and adapt to the GH first order system. Finally, in Section \ref{Sec:Results} we present our results where we consider several IBVPs involving gravitational wave scattering with boundaries placed inside and just outside of the horizons of spinning and non-spinning black holes.




\section{Covariant Summation By Parts}

\label{Sec:CovSBP}

A popular way to maintain stability in numerical hyperbolic IBVPs is to use finite differencing operators that satisfy summation by parts (SBP). Using SBP operators has been proven to be closely related to the well-posedness and the numerical stability of many conservative problems \cite{SBP_Review,Shifted_Wave}. The SBP property is usually discussed with respect to flat space and one dimension, defining a discrete analog of traditional integration by parts. Here, we make an effort to generalize the one dimensional covariant version of this property we developed in \cite{Dailey_2023} into a multidimensional approach that is applicable to spacetimes without symmetries.

Consider a scalar field $\phi$ that satisfies the massless Klein-Gordon equation $\nabla^\mu\nabla_\mu\phi=0$, where $\nabla_\mu$ is the covariant derivative with respect to the metric $g_{\mu\nu}$. This equation can be reduced to a first order system with the definition of two auxiliary variables: $\Pi\equiv-n^\mu\partial_\mu\phi$ and ${\psi_i\equiv\partial_i\phi}$, where $n^\mu$ is the timelike future-directed unit normal to the 3-space with metric $\gamma_{ij}$. In this work, Greek indices run over space and time, while Latin indices run over space only. If we consider a time independent background metric, we obtain the following system of evolution equations:
\begin{align}
\partial_t \phi&=\beta^i\psi_i -\alpha\Pi\,,\label{Eq.StaticSystem1}\\
\partial_t\psi_i+ \partial_i(\alpha\Pi-\beta^j\psi_j)&=0\,,\label{Eq.StaticSystem2}\\
\partial_t \Pi+\bar\nabla_i(\alpha\psi^i-\beta^i\Pi)&= 0\,,\label{Eq.StaticSystem3}
\end{align}
where $\bar\nabla_if^i=(\sqrt{\gamma})^{-1}\partial_i(\sqrt{\gamma}\,f^i)$ is the vector 3-divergence, $\gamma$ is the determinant of $\gamma_{ij}$, $\alpha$ is the lapse, and $\beta^i$ is the shift.  In this system, Eq.~(\ref{Eq.StaticSystem1}) is just the definition of $\Pi$, Eq.~(\ref{Eq.StaticSystem2}) enforces the commutativity of temporal and spatial derivatives, and Eq.~(\ref{Eq.StaticSystem3}) enforces the massless Klein-Gordon equation. The equations are written so that the left hand side contains the principal part, while the right hand side contains only source terms.

Since a time independent metric necessarily contains a timelike Killing vector $\xi^\mu$, the background spacetime is stationary and there exists a conserved current ${J^\mu=T^{\mu\nu}\xi_\nu}$, where $T_{\mu\nu}$ is the stress-energy tensor. If we define the stationary energy
\begin{align}
    E_s&=\frac{1}{2}\int\left[\alpha(\Pi^2+\psi_i\psi^i)-2\Pi\beta^i\psi_i\right]\,dV\,,
\end{align}
where $dV$ is the volume element of the 3-space, we can rewrite the conservation property $\nabla_\mu J^\mu=0$ to see that this energy is locally conserved:
\begin{align}
    \partial_tE_s= \oint(\alpha\Pi-\beta^j\psi_j) s_i(\alpha\psi^i-\beta^i\Pi)\,dS\,.
\end{align}
The right hand side describes the flux through a boundary surface $S$ and $s_i$ is the spacelike unit normal to that boundary. If we define the scalar $u\equiv\alpha\Pi-\beta^j\psi_j$ and the vector $v^i\equiv\alpha\psi^i-\beta^i\Pi$, we can write an ``integration by parts'' rule  with respect to this conservation law:
\begin{align}
    \int u\bar\nabla_iv^i\,dV+\int(\partial_i u)v^i\,dV=\oint u s_i v^i\,dS\,. \label{Eq:CovScalar}
\end{align}
If we instead define a vector field $F^i\equiv uv^i$, we see that this is nothing more than the covariant divergence theorem:
\begin{align}
    \int \bar\nabla_iF^i\,dV=\oint s_i F^i\,dS\,. 
\end{align}

Now we consider functions discretely sampled on a grid. A traditional one dimensional SBP derivative operator $D$, along with its norm operator $\Sigma$, satisfy
\begin{align}
u^{\rm T}\Sigma Dv + (D u)^{\rm T}\Sigma v=u^{\rm T}Bv\,,\label{Eq.reg_SBP}
\end{align}
where $B=\mathrm{diag}(-1,0,\dots,0,1)$,
for any two functions $u$ and $v$ sampled on a grid. In dimensions greater than one, $v$ will take the role of a vector valued function\footnote{The boundary operator has this form if grid values on the boundary are included. It has a different interpolating/extrapolating form if this is not the case (see for example \cite{MATTSSON2017255}).}. With a few alterations, the covariant version of this property in 3D can be written as
\begin{align}
u^\mathrm{T}W\sD_i v^i + (D_i u)^\mathrm{T}W v^i  = u^\mathrm{T} S_i v^i\,.\label{Eq.SBP}
\end{align}
Here, the operator $\sD_i \equiv \Gamma^{-1}D_i\Gamma$
approximates the 3-divergence, the operator $D_i$ approximates the scalar gradient, and $W\equiv\Sigma\sW$ is the covariant norm operator that approximates 3-volume integration, where operator $\sW$ is a diagonal matrix with the values of $\sqrt{\gamma}$ injected along that diagonal. Finally, ${S_i\equiv B \sS_i}$ is the boundary operator that approximates covariant surface integrals, where $\sS_i$ are diagonal matrices with the values of $s_i\sqrt{\sigma}$ injected along those diagonals and $\sigma$ is the determinant of the metric on the 2-surface of the boundary. This is to be directly compared with the continuous version Eq.~(\ref{Eq:CovScalar}). This is not the only way to think about the covariant SBP energy estimate; one might include an arbitrary positive definite symmetrizer for example, but we find this way particularly elegant. 

The SBP scheme we employ uses finite differencing operators that act on each of the variables $u$ and $v^i$, as is suggested by the left hand side of Eqs.~(\ref{Eq.StaticSystem2}) and (\ref{Eq.StaticSystem3}), which is in contrast to expanding the definition of these variables into multiple terms using the product rule of continuous calculus, as is common in numerical techniques. This is important for maintaining SBP, as discrete operators generally do not obey the product rule \cite{Mishra2010}. The evolution equations are then reminiscent of the flux-conservative form common in hydrodynamic systems.

The SBP formalism defined by Eq.~(\ref{Eq.SBP}) is completed by using an SBP operator in place of $D_i$. Although we will focus on finite differencing based SBP operators in this work, in principle SBP operators based on other methods, such as spectral methods, can complete this SBP formalism as well. The system of equations we used as an example in this section happens to have a local conservation law on which we can base the SBP property. This is not true in general for an arbitrary system of equations, like Einstein's equations. Instead, as long as the system of equations can be shown to be symmetric hyperbolic, there exists an energy estimate with respect to the principal part of the system \cite{Gundlach_2006}. It is this energy estimate that we aim our SBP method to preserve.\\


\section{The Generalized Harmonic First Order System}
\label{Sec:GenHarmonic}

As first introduced by \cite{Garfinkle2002,Szil_gyi_2003}, the Einstein field equations can be cast as a symmetric hyperbolic system by introducing a Generalized Harmonic (GH) gauge. Many successful numerical relativity simulations have since been based on this formulation (for example \cite{Pretorius_2005,Pretorius_2005_2,PhysRevD.107.064013}). 
The GH evolution equations can also be written as a symmetric hyperbolic first order system of equations \cite{Lindblom_2006}. To this end, a set of gauge functions $H_\mu$ are defined on a Lorentzian spacetime with metric $g_{\mu\nu}$ and coordinates $x^\mu$:
\begin{align}
H_\mu(x^\alpha)= -g_{\mu\sigma}\nabla_\nu\nabla^\nu x^\sigma = g^{\alpha\beta}\Gamma_{\mu\alpha\beta}\equiv \Gamma_\mu\,,
\end{align}
where $\Gamma_{\alpha\mu\nu}$ is the connection with respect to the metric $g_{\mu\nu}$. For $H_\mu=0$, this becomes the harmonic coordinate condition, but for an arbitrary set of spacetime functions, any gauge condition can be specified in principle. This set of gauge conditions can also be expressed as evolution equations for the lapse and shift:
\begin{align}
\partial_t\alpha-\beta^i\partial_i\alpha &= \alpha^2\left(n^\mu H_\mu-K\right)\,,\label{Eq:GaugeEvolution1}\\
\partial_t\beta^i-\beta^j\partial_j\beta^i &= \alpha^2\gamma^{ij}\left(\gamma^{kl}\Gamma_{jkl}-H_j-\partial_j\ln\alpha\right)\,,\label{Eq:GaugeEvolution2}
\end{align}
where $K=\nabla^\mu n_\mu$ is the trace of the extrinsic curvature $K_{\mu\nu}$. 

It has also been demonstrated that the evolution of $H_\mu$ can be achieved alongside the Einstein field equations \cite{Pretorius_2005,Pretorius_2005_2}, where a wave-like evolution scheme is adopted. The symmetric hyperbolicity of the GH evolution equations can be destroyed with an arbitrary evolution scheme for $H_\mu$. By picking an evolution scheme of second order, such as a wave operator (evolved as a first order system of equations) the symmetric hyperbolicity of the GH evolution equations is maintained, as the principal part is not altered. For simplicity, in this work we will focus on a function specification of $H_\mu(x^\nu)$, where we effectively use the gauge condition $\partial_t H_\mu=0$.

With the two auxiliary definitions $P_{\mu\nu}\equiv-n^\alpha\partial_\alpha g_{\mu\nu}$ and $d_{i\mu\nu}\equiv\partial_i g_{\mu\nu}$, we can write the symmetric hyperbolic GH evolution system as



\begin{widetext}
\begin{align}
\partial_t{g_{\mu\nu}} = &\; \beta^id_{i\mu\nu} - \alpha P_{\mu\nu} \,, \label{Eq:Evolution1}\\
\partial_t{d_{i \mu\nu}} + \partial_i(\alpha P_{\mu\nu}-\beta^j d_{j\mu\nu}) = &\;  0\,,\label{Eq:Evolution2} \\
\partial_tP_{\mu\nu}  + \bar\nabla_i (\alpha d\indices{^{\,i}_\mu_\nu}-\beta^iP_{\mu\nu}) =  &\;-(\partial_t\ln\sqrt{\gamma}) P_{\mu\nu}+ 2\alpha\partial_{(\mu}H_{\nu)}-2\alpha H_\sigma\Gamma\indices{^\sigma_\mu_\nu} -\alpha\Gamma^\sigma \partial_\sigma g_{\mu\nu}\nonumber\\&\;+2\alpha g^{\epsilon\sigma}g^{\lambda\rho}(\partial_\lambda g_{\epsilon\mu}\partial_\rho g_{\sigma\nu}-\Gamma_{\mu\epsilon\lambda}\Gamma_{\nu\sigma\rho})\,,\label{Eq:Evolution3}
\end{align}
\end{widetext}
where the operator $\bar\nabla_i$ acts still as a \emph{vector} divergence so that $\bar\nabla_if\indices{^i_\mu_\nu}\equiv(\sqrt{\gamma})^{-1}\partial_i(\sqrt{\gamma}\,f\indices{^i_\mu_\nu})$. In this system, Eq.~(\ref{Eq:Evolution1}) is just the definition of $P_{\mu\nu}$, Eq.~(\ref{Eq:Evolution2}) enforces the commutativity of temporal and spatial derivatives, and Eq.~(\ref{Eq:Evolution3}) enforces the vacuum Einstein field equations.

This system of first order equations has state vector ${\vec U = (g_{\mu\nu},d_{i\mu\nu},P_{\mu\nu})}$. It is understood that where Christoffel symbols, first-order metric derivatives, lapse, and shift appear in the evolution equations, they stand for elements of the state vector according to their definitions. For the Christoffel symbols we have
\begin{align}
2\Gamma_{\alpha\mu\nu}=\partial_\mu g_{\alpha\nu} + \partial_\nu g_{\alpha\mu} - \partial_\alpha g_{\mu\nu}\,,
\end{align}
and all first order metric derivatives are defined via the state vector with
\begin{align}
\partial_t g_{\mu\nu} &= \beta^i d_{i\mu\nu}-\alpha P_{\mu\nu}\,,\\
\partial_i g_{\mu\nu} &= d_{i\mu\nu}\,.
\end{align}
The time derivative of the 3-metric determinant is expressed as
\begin{align}
\partial_t \ln\sqrt{\gamma}=\frac{1}{2}\gamma^{ij}\partial_t\gamma_{ij}=\frac{1}{2}\gamma^{ij}(\beta^k d_{kij}-\alpha P_{ij})\,,
\end{align}
and the lapse and shift are set via the inverse metric components
\begin{align}
\alpha^2 = (-g^{tt})^{-1}\,,\quad
\beta^i = \alpha^2 g^{ti}\,.
\end{align}
This system is very similar to that defined originally by \cite{Lindblom_2006}, except we have altered the principal part in the same sense as suggested by Section \ref{Sec:CovSBP}.

Since the first order GH formulation is symmetric hyperbolic, it admits an energy estimate with respect to the principal part of the system \cite{Gundlach_2006, Lindblom_2006}:
\begin{align}
E_g=\frac{1}{2}\int m^{\mu\nu}m^{\epsilon\sigma}\Big[&\alpha(P_{\mu\epsilon}P_{\nu\sigma}+d\indices{^i_\mu_\epsilon}d_{i\nu\sigma})\nonumber\\&-2\beta^i P_{\mu\epsilon}d_{i\nu\sigma}\Big]\,dV\,,
\end{align}
where $m^{\mu\nu}$ is any positive definite metric (we use ${m^{\mu\nu}=n^\mu n^\nu + \gamma^{\mu\nu}}$). This energy is  locally conserved
\begin{align}
\partial_tE_g\sim \oint m^{\mu\nu}m^{\epsilon\sigma}&(\alpha P_{\mu\epsilon}-\beta^j d_{j\mu\epsilon})\nonumber\\& \times s_i(\alpha d\indices{^i_{\nu\sigma}}-\beta^i P_{\nu\sigma})\,dS\, ,
\end{align}
if we ignore the source terms on the right hand side of GH evolution equations. By writing these equations as we have, and defining ${u_{\mu\nu}\equiv\alpha P_{\mu\nu}-\beta^j d_{j\mu\nu}}$ and ${v\indices{^i_\mu_\nu}\equiv\alpha d\indices{^i_\mu_\nu}-\beta^i P_{\mu\nu}}$, we can write an ``integration by parts'' rule for this system, keeping only terms from the principal part:
\begin{align}
    \int m^{\mu\nu}m^{\epsilon\sigma}&\left[u_{\mu\epsilon}\bar\nabla_iv\indices{^i_\nu_\sigma}+(\partial_i u_{\mu\epsilon})v\indices{^i_\nu_\sigma}\right]\,dV\nonumber\\&\sim\oint m^{\mu\nu}m^{\epsilon\sigma}u_{\mu\epsilon} s_i v\indices{^i_\nu_\sigma}\,dS\,. 
\end{align}
Just as in Section~(\ref{Sec:CovSBP}), we endeavor to write a discrete analog of this property. We sample the functions $u_{\mu\nu}$ and $v\indices{^i_\mu_\nu}$ to a grid and obtain
\begin{align}
(u_{\mu\nu})^\mathrm{T}W^{\mu\nu\alpha\beta}\sD_i v\indices{^i_\alpha_\beta}& + (D_i u_{\mu\nu})^\mathrm{T}W^{\mu\nu\alpha\beta} v\indices{^i_\alpha_\beta} \nonumber\\&
= (u_{\mu\nu})^\mathrm{T} S\indices{_i^\mu^\nu^\alpha^\beta} v\indices{^i_\alpha_\beta}\,.
\end{align}
The operators $\sD_i$ and $D_i$ act the same as in Section~(\ref{Sec:CovSBP}), but the covariant norm operator is ${W^{\mu\nu\alpha\beta}\equiv\Sigma\sW^{\mu\nu\alpha\beta}}$ where ${\sW^{\mu\nu\alpha\beta}}$ is a diagonal matrix with the values of $\sqrt{\gamma}m^{\mu\alpha}m^{\nu\beta}$ injected along that diagonal. Finally, ${S\indices{_i^\mu^\nu^\alpha^\beta}=B \sS\indices{_i^\mu^\nu^\alpha^\beta}}$ is the boundary operator for this system, where $\sS\indices{_i^\mu^\nu^\alpha^\beta}$ are diagonal matrices with the values of $s_i\sqrt{\sigma}m^{\mu\alpha}m^{\nu\beta}$ injected along those diagonals. This allows for the direct application of SBP methods to this system of evolution equations if the substitution $\bar\nabla_i\rightarrow \sD_i$ and $\partial_i\rightarrow D_i$ is made with an appropriate SBP derivative operator.

The GH formulation introduces the set of gauge constraints ${\sC_\alpha \equiv g^{\mu\nu}\Gamma_{\alpha\mu\nu}-H_{\alpha}}$ and the first order reduction introduces the set of spatial derivative constraints ${\sC_{i\mu\nu}\equiv D_ig_{\mu\nu}-d_{i\mu\nu}}$. To ensure that these constraints do not grow during the evolution, constraint damping terms are added to the right hand side of the evolution equations \cite{Lindblom_2006}:
\begin{align}
\partial_t{d_{i \mu\nu}} = &\; \cdots + \gamma_2 \alpha\sC_{i\mu\nu}\,, \\
\partial_tP_{\mu\nu} = &\; \cdots -\gamma_2\beta^i\sC_{i\mu\nu}+ \gamma_0\alpha(2n_{(\mu}\sC_{\nu)}-g_{\mu\nu}n^\epsilon\sC_\epsilon)\,,
\end{align}
which results in exponential damping of constraint violations as long as the rate parameters $\gamma_{0,2}$ are positive. The time derivatives of the gauge constraints ($\partial_t\sC_\alpha$) can be shown to effectively replace the Einstein constraints $\sM^\nu=n_\mu G^{\mu\nu}$ \cite{Lindblom_2006,Brown_2011}, where $G^{\mu\nu}$ is the Einstein tensor, which allows us to infer that $\sM^\mu=0$ from static behavior in $\sC^\mu$. In fact, if one can show $\sC_\mu=\partial_i\sC_\mu=0$, then $\partial_t\sC_\mu=2\alpha\sM_\mu$.

\section{Initial Conditions}
\label{Sec:ICs}

Given a completely arbitrary metric on the initial hypersurface $g_{\mu\nu}(t=0)=g^{\rm init}_{\mu\nu}$, we aim to set up initial conditions for the rest of the state-vector while controlling the set of constraints defined above. We can easily control the initial values of the constraints $\sC_{i\mu\nu}$ by setting
\begin{align}
d^{\rm init}_{i\mu\nu}=D_i g_{\mu\nu} -\sC^{\rm init}_{i\mu\nu}\,,
\end{align}
for an arbitrary set of functions $\sC^{\rm init}_{i\mu\nu}$. Next, we can control the initial values of $\sC_\alpha$ by setting 4 of the initial components of $P_{\mu\nu}$. The definition of $\sC_{\alpha}$ gives 
\begin{align}
\sC_\alpha&=\left(n^{(\mu}\delta\indices{^{\nu)}_\alpha}-\frac{1}{2}g^{\mu\nu}n_\alpha\right)P_{\mu\nu}\nonumber\\&+\left(\gamma^{i(\mu}\delta\indices{^{\nu)}_\alpha}-\frac{1}{2}g^{\mu\nu}\gamma\indices{^i_\alpha}\right)d_{i\mu\nu}-H_{\alpha}\,.
\end{align}
We then define two projection operators that transfer between components of symmetric second rank tensors:
\begin{align}
   \delta\indices{_{(\mu}^{(\alpha}}\delta\indices{_{\nu)}^{\beta)}} = A\indices{_\mu_\nu^\alpha^\beta} + B\indices{_\mu_\nu^\alpha^\beta}\,.
\end{align}
The operator $A\indices{_\mu_\nu^\alpha^\beta}$ acts to pick out the 4 degrees of freedom in $P_{\mu\nu}$ responsible for the values of $\sC_\alpha$:
\begin{align}
A\indices{_\mu_\nu^\alpha^\beta}\equiv n_{\mu}n_{\nu}n^\alpha n^\beta-2n_{(\mu}\gamma\indices{_{\nu)}^{(\alpha}}n^{\beta)}+n_{\mu}n_{\nu}\gamma^{\alpha\beta}\,,
\end{align}
which has a complimentary operator
\begin{align}
   A\indices{_\mu_\nu^\alpha} \equiv n_{(\mu}\gamma\indices{_{\nu)}^\alpha}-n_\mu n_\nu n^\alpha\,,
\end{align}
that satisfies $A\indices{_\mu_\nu^\alpha^\beta}A\indices{_\alpha_\beta^\epsilon}=A\indices{_\mu_\nu^\epsilon}$. Together, they complete the relation
\begin{align}
   A\indices{_\mu_\nu^\alpha^\beta}P^{\rm init}_{\alpha\beta} = A\indices{_\mu_\nu^\epsilon}\big(&2\gamma^{i\alpha}d_{i\alpha\epsilon}-g^{\alpha\beta}\gamma\indices{^i_\epsilon}d_{i\alpha\beta}\nonumber\\&-2H_\epsilon-2\sC_\epsilon^{\rm init}\big)\,,
\end{align}
for an arbitrary set of functions $\sC_\epsilon^{\rm init}(x^i)$. The other components of $P^{\rm init}_{\mu\nu}$ are arbitrary, and can be set however desired with the operator
\begin{align}
B\indices{_\mu_\nu^\alpha^\beta}\equiv \gamma\indices{_{(\mu}^{(\alpha}}\gamma\indices{_{\nu)}^{\beta)}}-n_{\mu}n_{\nu}\gamma^{\alpha\beta}\,.
\end{align}
In principle, these remaining components of $P_{\mu\nu}$ selected by this operator can be assigned meaning as we will do in a similar way in the next section, but we don't find this necessary for this work. Our choices for the projection operators used here might not be unique.

For most practical applications, the choice to satisfy the constraints on the initial hypersurface (i.e. ${\sC^{\rm init}_\alpha=\sC^{\rm init}_{i\mu\nu}=0}$) will be desired, but allowing them to be set arbitrarily is very useful for numerical tests. Setting the values of $\sC^{\rm init}_\alpha=0$ does not necessarily imply that the Einstein constraints are satisfied however, as they are effectively controlled with $\partial_t\sC^{\rm init}_\alpha$, but with the addition of an appropriate rate of constraint damping, any initial value of $\partial_t\sC^{\rm init}_\alpha$ will vanish quickly.

\section{Boundary Conditions}
\label{Sec:BCs}

The generic method of applying BCs to systems of hyperbolic IBVPs is the method of characteristics. The characteristic modes and speeds are defined based on the eigenvectors and eigenvalues of the principal part of the system (see \cite{Alcubierre:1138167} for a review). Given a boundary hypersurface with purely spatial unit normal $s^\mu$ orthogonal to $n^\mu$ pointing out of the computational domain, the characteristic modes for this system are given by \footnote{Technically, with the addition of the derivative constraint damping, the definitions of $U^\pm_{\mu\nu}$ should contain an additional term $\gamma_2g_{\mu\nu}$, but for simplicity, we considered a characteristic decomposition based on first derivatives of the metric only, and we found that this works well. The formulas presented in this work can be cast in terms of the proper characteristics $\tilde U^\pm_{\mu\nu}$ with the relationship $\tilde U^\pm_{\mu\nu}=U^\pm_{\mu\nu}+\gamma_2 g_{\mu\nu}/\sqrt{2}$.}
\begin{align}
U^+_{\mu\nu} &\equiv k^\alpha\partial_\alpha g_{\mu\nu}=-\frac{1}{\sqrt{2}}(P_{\mu\nu}+s^id_{i\mu\nu})\,,\\
U^-_{\mu\nu} &\equiv \ell^\alpha\partial_\alpha g_{\mu\nu}=-\frac{1}{\sqrt{2}}(P_{\mu\nu}-s^id_{i\mu\nu})\,,\\
U^\perp_{\alpha\mu\nu} &\equiv\sigma\indices{_\alpha^\beta}\partial_\beta g_{\mu\nu}= \sigma\indices{_\alpha^i}d_{i\mu\nu}\,.
\end{align}
The characteristic modes are only unique up to a constant multiplicative factor, and we find it convenient to normalize them such that they are simply the derivatives in the direction of the outgoing null vector ${\ell^\mu=(n^\mu+s^\mu)/\sqrt{2}}$, the ingoing null vector ${k^\mu=(n^\mu-s^\mu)/\sqrt{2}}$, and the metric on the 2-surface of the boundary ${\sigma_{\mu\nu}=g_{\mu\nu}+2\ell_{(\mu}k_{\nu)}}$. The characteristics $U^\pm_{\mu\nu}$ travel along the light cone with coordinate speed $c_\pm=\pm\alpha-s_i\beta^i$. The characteristics $U^\perp_{\alpha\mu\nu}$ travel along $n^\mu$ with coordinate speed $c_\perp=-s_i\beta^i$. The method of characteristics dictates that all of the modes with a characteristic speed negative in value (i.e. propagating into the computational domain) require BCs.

The characteristic $U^-_{\mu\nu}$ at a point on the boundary is always incoming unless the boundary worldtube is not timelike there (such as if it is within an apparent horizon), so we assign to it the physical incoming degrees of freedom, and we will replace these components with our BCs (i.e. ${U^-_{\mu\nu}\rightarrow U^{-\rm BC}_{\mu\nu}}$). If the characteristic $U^\perp_{\alpha\mu\nu}$ is incoming, we set it by demanding the vanishing of the derivative constraints $\sC_{i\mu\nu}=0$. If $U^{-}_{\mu\nu}$ is the only incoming characteristic for example, the values of the state vector can be reconstructed on the boundary as
\begin{align}
P^{\rm BC}_{\mu\nu}&=-\frac{1}{\sqrt{2}}\big(U^+_{\mu\nu}+U^{-\rm BC}_{\mu\nu}\big)\,,\\
d^{\rm BC}_{i\mu\nu}&=U^\perp_{i\mu\nu}-\ell_i U^+_{\mu\nu}-k_{i}U^{-\rm BC}_{\mu\nu}\,.
\end{align}
We still need to understand which of the components of $U^{-\rm BC}_{\mu\nu}$ affect which degrees of freedom of the metric itself. Here, we follow and generalize the component projection formalism discussed in \cite{Lindblom_2006}.

We introduce the following decomposition of the identity operator that transfers between components of symmetric second rank tensors:
\begin{align}
   \delta\indices{_{(\mu}^{(\alpha}}\delta\indices{_{\nu)}^{\beta)}} = C\indices{_\mu_\nu^\alpha^\beta} + P\indices{_\mu_\nu^\alpha^\beta} + G\indices{_\mu_\nu^\alpha^\beta}\,.\label{Eq.decomp}
\end{align}
Each of these operators has a clear interpretation as to what incoming degrees of freedom in the metric they control, which are elaborated in the following subsections.

\subsection{Constraint Controlling BCs}

Four of the components of $U^{-\rm BC}_{\mu\nu}$ can be used to control the values of the gauge constraints $\sC_\alpha$ on the boundary. The operator
\begin{align}
   C\indices{_\mu_\nu^\alpha^\beta} \equiv \frac{1}{2}\sigma_{\mu\nu}\sigma^{\alpha\beta}-2\ell_{(\mu}\sigma_{\nu)}^{\;\;\;(\alpha}k^{\beta)}+\ell_\mu\ell_\nu k^\alpha k^\beta\,,
\end{align}
picks out these components. There is also a complementary operator
\begin{align}
   C\indices{_\mu_\nu^\alpha} \equiv \ell_{(\mu}\sigma\indices{_{\nu)}^\alpha}-\frac{1}{2}\sigma_{\mu\nu}\ell^\alpha-\frac{1}{2}\ell_\mu\ell_\nu k^\alpha\,,
\end{align}
that satisfies $C\indices{_\mu_\nu^\alpha^\beta}C\indices{_\alpha_\beta^\epsilon}=C\indices{_\mu_\nu^\epsilon}$. The form of these operators (originally written in \cite{Lindblom_2006}), comes directly from the  definition of the gauge constraints (${\sC_\alpha=g^{\mu\nu}\Gamma_{\alpha\mu\nu}-H_\alpha}$). This definition is equivalent to
\begin{align}
   C\indices{_\mu_\nu^\alpha^\beta}U^{-\rm BC}_{\alpha\beta}  = C\indices{_\mu_\nu^\epsilon}\Big(&2\ell^{\alpha}U^+_{\alpha\epsilon}-\ell_\epsilon g^{\alpha\beta}U^+_{\alpha\beta}+g^{\alpha\beta}U^\perp_{\epsilon\alpha\beta}\nonumber\\&-2g^{\alpha\beta}U^\perp_{\alpha\beta\epsilon} + 2H_\epsilon + 2\sC_\epsilon^{\rm BC}\Big)\,,
\end{align}
where we have allowed for some arbitrary functions $\sC_\mu^{\rm BC}(x^\alpha)$. In this fashion, the values of $\sC_\alpha$ on the boundary are specified by the BCs desired.

\subsection{Gravitational Wave Controlling BCs}

The operator $P\indices{_\mu_\nu^\alpha^\beta}$ is the traceless part of the projection onto the $2$ dimensional manifold of the boundary
\begin{align}
   P\indices{_\mu_\nu^\alpha^\beta} \equiv \sigma\indices{_{(\mu}^{(\alpha}}\sigma\indices{_{\nu)}^{\beta)}} - \frac{1}{2}\sigma_{\mu\nu}\sigma^{\alpha\beta} \,,
\end{align}
and controls the $2$ degrees of freedom that dictate physical incoming GWs \cite{Lindblom_2006}. We can inject a GW model through the boundary by picking an arbitrary model $h^{\rm BC}_{\mu\nu}$ and setting it with
\begin{align}
   P\indices{_\mu_\nu^\alpha^\beta}U^{-\rm BC}_{\alpha\beta} = P\indices{_\mu_\nu^\alpha^\beta}h^{\rm BC}_{\alpha\beta}\,.
\end{align}
It was shown in \cite{PhysRevD.71.064020} that the object $P\indices{_\mu_\nu^\alpha^\beta}s^i\partial_iU^-_{\alpha\beta}$ is related to the complex Newman-Penrose curvature scalar $\Psi_0$ ($\Psi_4$) at the outer (inner) boundary. The model $h^{\rm BC}_{\alpha\beta}$ can then be thought of as related to a line integral over $\Psi_{0}$ or $\Psi_4$ in the direction $s^i$ via the gradient theorem. If we decompose the 2-metric into two spatial directions using a set of orthonormal basis vectors $p^\mu$ and $q^\mu$ such that $\sigma_{\mu\nu}=p_\mu p_\nu+q_\mu q_\nu$, we can see (as expected) that there are only two components of this projection, and they are related to derivatives of $\sigma_{\mu\nu}$:
\begin{align}
   (p^\mu p^\nu-q^\mu q^\nu)P\indices{_\mu_\nu^\alpha^\beta}U^{-}_{\alpha\beta} &= (p^\mu p^\nu-q^\mu q^\nu)\ell^\alpha\partial_\alpha\sigma_{\mu\nu}\,,\\
   p^\mu q^\nu P\indices{_\mu_\nu^\alpha^\beta}U^{-}_{\alpha\beta} &= p^\mu q^\nu\ell^\alpha\partial_\alpha\sigma_{\mu\nu}\,.
\end{align}
The former controls the $+$ polarization and the latter controls the $\times$ polarization with respect to these basis vectors. 

In principle these components can also be set using an arbitrary function of the outgoing fields
\begin{align}
   h^{\rm BC}_{\alpha\beta} = h^{\rm BC}_{\alpha\beta}(U^+_{\mu\nu},U^\perp_{\sigma\mu\nu})\,,
\end{align}
but careful analysis and numerical testing is needed to understand what a specific choice implies about the GWs that would be injected into the computational domain in response to the GWs incident on the boundary itself. One way to do this is to use a conservation law to gain some insight.

Since in 3D general relativity does not have local conservation laws, we can consider instead a class of quasi-local conservation laws \cite{McGrath_2012}. These are called quasi-local because instead of relating the change in a volume integral to the flux through a 2-surface, they relate the change in a 2-surface integral to the flux through a 2-surface, and hence they only depend on the geometry in the neighborhood of the 2-surface. Just as we have done in \cite{Dailey_2023}, we can in principle define BCs on GWs that control the flux portion of a quasi-local conservation law to define what it means to reflect gravitational radiation. We plan to refine this idea in future work.


\subsection{Gauge Controlling BCs}

Finally, the operator
\begin{align}
   G\indices{_\mu_\nu^\alpha^\beta} \equiv 2\ell_{(\mu}k_{\nu)}k^{(\alpha}\ell^{\beta)}-2k_{(\mu}\sigma\indices{_{\nu)}^{(\alpha}}\ell^{\beta)}+k_{\mu}k_{\nu}\ell^{\alpha}\ell^{\beta}\,,
\end{align}
controls the 4 incoming gauge degrees of freedom. It has a complementary operator
\begin{align}
   G\indices{_\mu_\nu^\alpha} \equiv k_{(\mu}\sigma\indices{_{\nu)}^\alpha}+k_{\mu}k_{\nu}\ell^\alpha+\ell_{(\mu}k_{\nu)}k^\alpha\,,
\end{align}
that satisfies $G\indices{_\mu_\nu^\alpha^\beta}G\indices{_\alpha_\beta^\epsilon}=G\indices{_\mu_\nu^\epsilon}$. One can relate the remaining components of $U^{-\rm BC}_{\mu\nu}$ to the incoming derivatives of $n^\mu$, effectively controlling the incoming derivatives of the lapse and shift
\begin{align}
   \ell^\sigma\partial_\sigma n^\beta=-\left(g^{\beta\nu}+\frac{1}{2}n^{\beta} n^\nu\right)n^\mu U^-_{\mu\nu}\,.
\end{align}




Although it was shown in Eqns.~(\ref{Eq:GaugeEvolution1}-\ref{Eq:GaugeEvolution2}) that the selection of $H_\mu$ can be thought of as first order evolution equations for the lapse and shift, these equations are \emph{implicit} as they are not directly evolved, and are implied instead by the second order system of GH evolution equations. This means that we are free to choose a set of first order evolution equations for the lapse and shift only on the boundaries, which is what the operator $G\indices{_\mu_\nu^\alpha^\beta}$ specifies. For simplicity we can select a model $\sG^{\rm BC}_{\alpha\beta}$ and apply it with
\begin{align}
   G\indices{_\mu_\nu^\alpha^\beta}U^{-\rm BC}_{\alpha\beta} = G\indices{_\mu_\nu^\alpha^\beta}\sG^{\rm BC}_{\alpha\beta}\,.
\end{align}


\subsection{Discussion on BC choices}

Together with a user specified set $\sC^{\rm BC}_\mu$, $h^{\rm BC}_{\mu\nu}$ , and $\sG^{\rm BC}_{\mu\nu}$, we fix all 10 incoming degrees of freedom in the metric, each component of which has a clear interpretation. 

A natural choice for the value of the constraints on the boundary is $\sC^{\rm BC}_\mu=0$. One can also inject a model of constraint violating modes for numerical tests and constraint damping verification if desired. Although $h^{\rm BC}_{\mu\nu}$ and $\sG^{\rm BC}_{\mu\nu}$ are left arbitrary, bad choices can be made, such as choices that introduce discontinuities into the metric. The GH equations can develop shocks when discontinuous initial data and BCs are used. The simplest choice that avoids this possibility is $h^{\rm BC}_{\mu\nu}=\sG^{\rm BC}_{\mu\nu}=U^{-\rm init}_{\mu\nu}$, which is effectively the same BCs used for these components by default in \cite{Lindblom_2006}. We will use this choice for simplicity in this work.


It is important to note that the BCs here are applied via first derivatives of the metric, effectively acting as a Dirichlet BC with respect to the evolved state-vector, whereas in \cite{Lindblom_2006} BCs are applied via \emph{second} derivatives of the metric, effectively controlling $\partial_t U^-_{\mu\nu}$ rather than $U^-_{\mu\nu}$. The control of $\partial_t U^-_{\mu\nu}$ allows for the imposition of constraint transparent/absorbing boundaries, where constraint violating modes are allowed to leave the computational domain through the boundaries. In contrast, the BCs used here enforce the \emph{value} of $\sC_\mu$, which was first shown to be well-posed in \cite{Babiuc_2007} and tested numerically in \cite{Rinne_2007}, although this BC method is not in common use today. These BCs thus act like a Dirichlet condition on $\sC_\mu$, allowing constraint violating modes to reflect off of the boundaries of the domain. We consider this behavior acceptable as long as the constraint damping rate is adequate. The reason for this difference is so that the BCs can be implemented using SATs, a weak application of BCs that has been shown to be essential in ensuring SBP properties are obeyed properly \cite{SBP_Review}. In principle, one could evolve an equation for $\partial_t\sC^{\rm BC}_\alpha$ alongside the GH evolution equations and still apply the BCs as we do here to impose the same constraint transparent boundaries of \cite{Lindblom_2006}, but we do not find this necessary. In fact, by imposing $\sC^{\rm BC}_\alpha=\partial_t\sC^{\rm BC}_\alpha=0$, we ensure that in the vicinity of the boundary, the Einstein constraints are satisfied, which is desirable in the strong gravity region of the spacetime where constraint violations and numerical errors are likely to spoil the evolution.



\section{Embedded Boundary SBP Methods}
\label{Sec:EmbeddedMethods}

In an effort to use the same modern numerical methods that we have used in our previous work, we consider here a class of embedded boundary SBP methods. This relatively recently introduced class of numerical methods allow for the evolution of systems on arbitrarily shaped domains. The general idea is to use a rectangular grid, and Cartesian-like coordinates, to evolve the equations in the bulk, where spatial derivatives are calculated using an SBP based finite differencing technique whose stencils terminate wherever they intersect a boundary of the domain.

One such method introduced in \cite{MATTSSON2017255} allows the boundary position to be between grid points, and the boundary values are obtained and BCs applied using interpolation/extrapolation operators. This method is easy to implement, as nothing more than a rectangular grid of values is needed, with points inside a closed boundary within this bulk grid just ``turned off'' as they are not evolved. The authors of \cite{MATTSSON2017255} do outline several complications of this method that require extra care, including special situations where thin domain geometries do not allow for enough grid points to fit a finite differencing stencil.

\begin{figure}[t]
\includegraphics[width=0.7\linewidth]{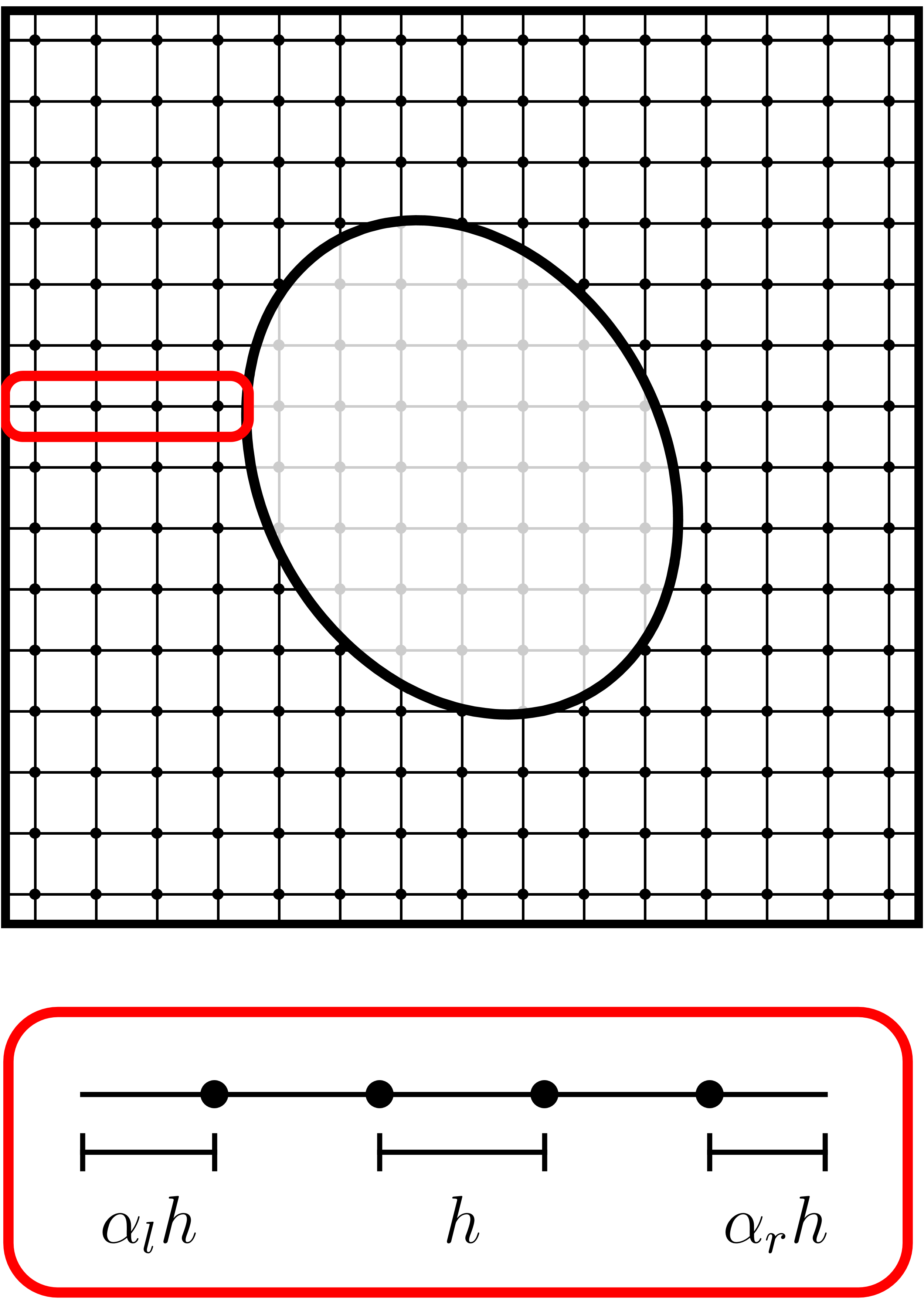}
\caption{Visualization of the embedded boundary method from \cite{MATTSSON2017255} on a 2D slice, here shown with a square outer boundary and an elliptical inner boundary (bold lines/curves), with the computational grid points shown connected with thin lines. The light grey grid points within the ellipse exist in memory on the bulk grid, but are not evolved. The embedded boundary method considers all of the connecting grid lines and defines finite differencing stencils along these lines that depend on the grid spacing $h$ as well as the distance parameters $\alpha_l$ and $\alpha_r$ that dictate where the boundary of the domain lies relative to the ends of each of the grid lines. The red region zooms in to one such line to visualize these parameters.}\label{Fig:GridDemo}
\end{figure}

\begin{figure*}[t]
\centering
Schwarzschild black hole excised inside of the horizon\par\medskip
\includegraphics[width=\textwidth]{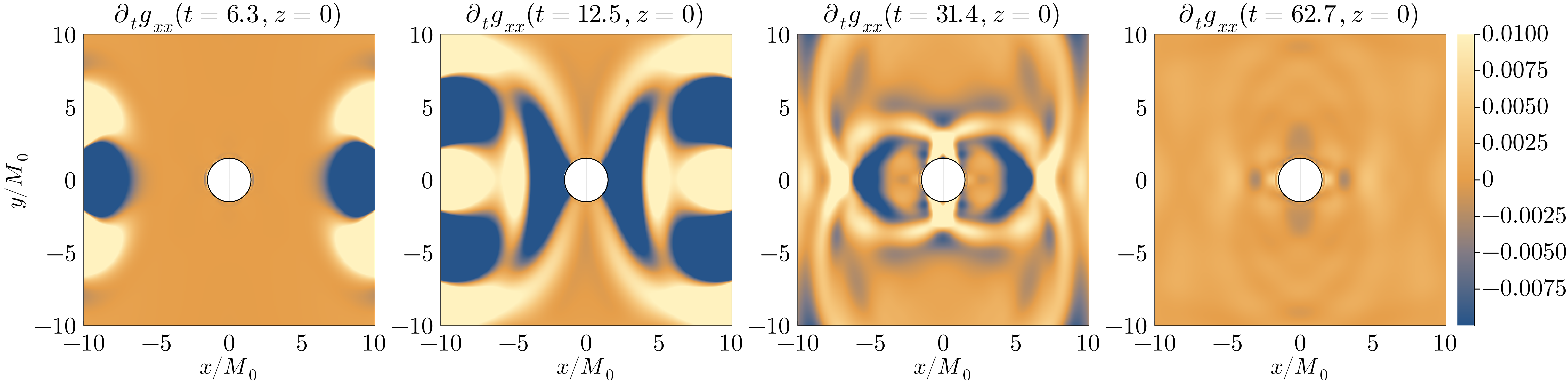}
Fractional Change in Quasi-Local Measures\par
\includegraphics[width=0.95\textwidth]{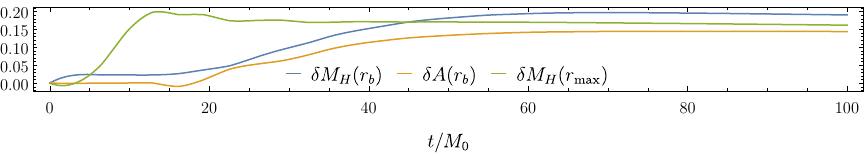}
\caption{Simulation of GW scattering on a Schwarzschild black hole of initial mass $M_0=1$. The black hole is excised \emph{inside} the horizon at $r_b=1.5M_0$. Top panels show $\partial_t g_{xx}$ at several times on the $z=0$ slice. Bottom panel shows the fractional change in the surface area of the inner boundary $A(r_b)$ as well as the Hawking mass $M_H$ evaluated at $r_b$ and the largest sphere in the cubical domain $r_\mathrm{max}$.}\label{Fig:NoSpin}
\end{figure*}

Another method introduced in \cite{SHARAN2022111341} instead includes grid values on the boundary itself. For every point where a connecting grid line intersects a boundary, a grid value located there is introduced and evolved. This method is less straightforward to implement, as the points on a boundary do not necessarily fit in the normal rectangular structure of a grid array, and have to be stored separately and communicate with the bulk domain properly. Adding to the complexity, in order for the points on a boundary to be evolved, one also needs to interpolate finite differencing stencils along directions orthogonal to the intersecting stencil line, as there is not necessarily a stencil possible in these directions with the existing rectangular grid values. However, this method avoids many of the complications of thin domain geometries from the previous method.

As far as we are aware, an embedded boundary SBP finite differencing method that has been proven for dimensions more than one does not yet exist. Both of the methods we cite prove SBP for one dimension, then construct a multidimensional approach as if each connecting line in the grid is its own one dimensional problem. This falls short of an SBP proof for multidimensional problems, and \cite{MATTSSON2017255} adds numerical dissipation to the system to ensure stability. The authors from \cite{SHARAN2022111341} report to solve nonlinear problems without the use of numerical dissipation however.

\begin{figure*}[t]
\centering
Kerr black hole excised inside of the horizon\par\medskip
\includegraphics[width=\textwidth]{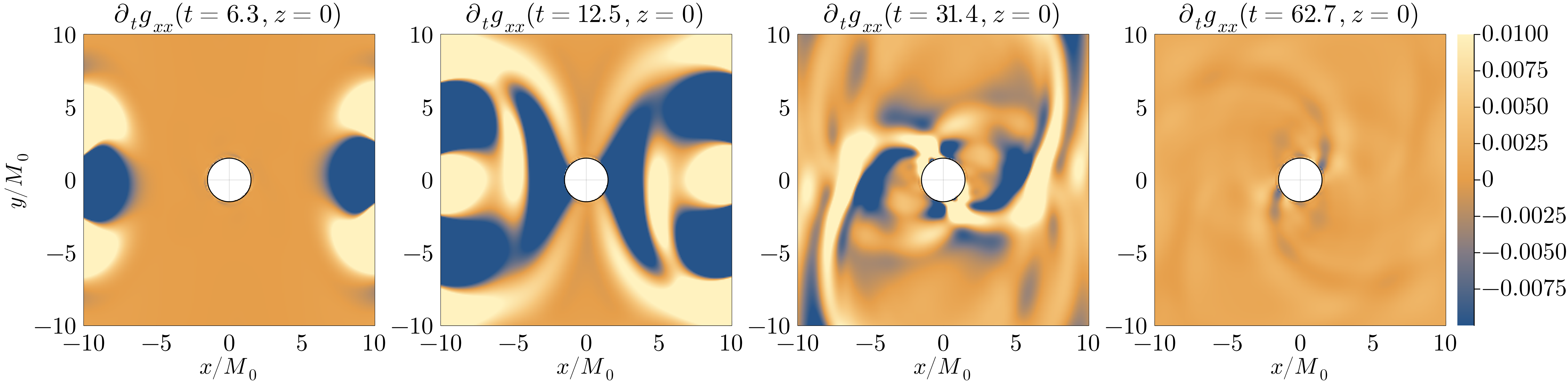}
Fractional Change in Quasi-Local Measures\par
\includegraphics[width=0.94\textwidth]{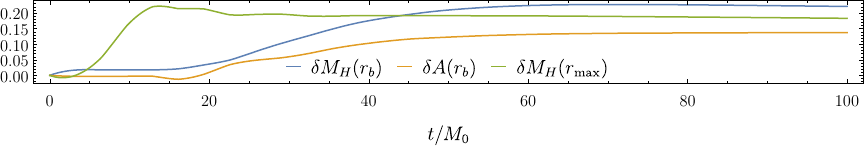}
\caption{Simulation of GW scattering on a Kerr black hole of initial mass $M_0=1$ and spin parameter $a=0.5M_0$.
The black hole is excised \emph{inside} the horizon at $r_b=1.5M_0$. Top panels show $\partial_t g_{xx}$ at several times on the $z=0$ slice. Bottom panel shows the fractional change in the surface area of the inner boundary $A(r_b)$ as well as the Hawking mass $M_H$ evaluated at $r_b$ and the largest sphere in the cubical domain $r_\mathrm{max}$.}\label{Fig:Spin}
\end{figure*}

\begin{figure}[h]
\includegraphics[width=\linewidth]{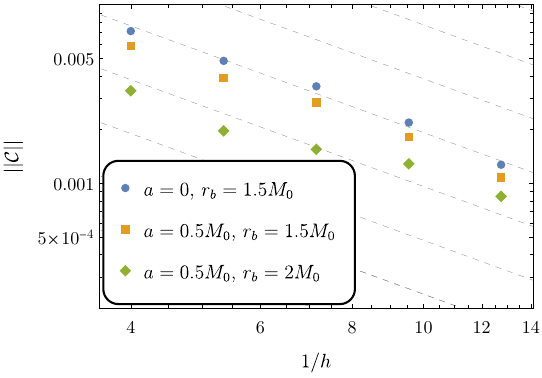}
\caption{Convergence of the constraint norm evaluated at $t=100M_0$ as the resolution is increased for the various simulations presented. Dashed lines are $\propto h^{1.5}$ to make it apparent that since the operators we use are second order accurate in the interior and first order at the boundaries, a global convergence rate of about $1.5$ is achieved.}\label{Fig:Cons}
\end{figure}

In this work, we adopt the method from \cite{MATTSSON2017255}. The operators from this method are defined in terms of two distance parameters, $\alpha_l$ and $\alpha_r$, that designate where the physical boundary is (in units of the grid spacing) from its closest grid points on the left and right of the stencil respectively, which is visualized in Figure \ref{Fig:GridDemo}. These parameters were originally defined to range from $-0.5\leq\alpha_{l,r}<0.5$, positive (negative) values indicating that the last grid point is just inside (outside) of the boundary relative to the computational domain. This is inconvenient however, as this parameter range allows for trapped points, grid points that cannot be assigned a finite differencing stencil in one or more directions, and thus the value of the derivative must be extrapolated there. This can be avoided completely (specifically for the case of a convex closed boundary with the computational domain surrounding it) if the parameter range is restricted to $0\leq\alpha_{l,r}<1$ instead. For the fourth order and sixth order accurate stencils in \cite{MATTSSON2017255}, the SBP norm is not positive definite for this parameter range, but the second order accurate operators $D_{2-1}$ are, so for simplicity we will use these operators in this work with this parameter range. This ensures that the boundary is always outside of the computational grid array, and the values on the boundary are always extrapolated rather than interpolated. As is required for SBP operators with diagonal norms, there is reduced accuracy near the boundary points. The operator $D_{2-1}$ is second order accurate in the interior but only first order accurate near the boundaries.

From \cite{MATTSSON2017255}, we obtain the SBP derivative operator $D_{2-1}(\alpha_{l,r})$, its corresponding norm operator $\Sigma(\alpha_{l,r})$, and a set of first order accurate extrapolation operators $e_{l,r}(\alpha_{l,r})$. Since the boundary does not necessarily lie on a grid point, the application of BCs needs to be extrapolated. At a boundary intersecting a grid line on its left for example, the SAT application of BCs is
\begin{align}
    \partial_t P_{\mu\nu} &= \cdots - c_-\Sigma^{-1}e_l\left[P^{\rm BC}_{\mu\nu}-e_l\, P_{\mu\nu}\right]\,,\\\partial_t d_{i\mu\nu}&= \cdots -c_-\Sigma^{-1}e_l\left[d^{\rm BC}_{i\mu\nu}-e_l\,d_{i\mu\nu}\right]\,.
\end{align}
This is very similar to adding an exponential decay term to the evolution equations at the boundary (as is the general idea of SAT techniques), except that the value of the state vector at the boundary is extrapolated, and the exponential decay term is applied at the boundary point using an extrapolation operator as well. The decay rate is set to the incoming coordinate speed $c_-$ as we showed is necessary in \cite{Dailey_2023}, and thus the BCs are applied more weakly the closer the boundary worldtube is to a null hypersurface where $c_-\rightarrow 0$.

\section{Results}
\label{Sec:Results}
Here, we demonstrate several 3D simulations to showcase our altered formulation, our new BC framework, and the embedded boundary numerical method. We study the evolution of GW scattering on spinning and non-spinning black holes and cases where the center of the hole is excised just inside or just outside of the apparent horizon.

In the following set of simulations presented, we keep as many things the same as possible to aid in direct comparison. We use initial mass $M_0=1$ for the black holes and constraint damping parameters $\gamma_0=\gamma_2=1/M_0$. We also add  numerical dissipation to aid in the stability of the simulation. This is done as
\begin{align}
    \partial_t\vec U=\cdots + \varepsilon D_4 \vec U\,,
\end{align}
where $\varepsilon$ is the amount of dissipation and the dissipation operator $D_4$ that pairs with the second order finite differencing operators we use in this work is found in \cite{MATTSSON2017255}. We use $\varepsilon = 0.1$ in all of the simulations presented here.

\begin{figure*}[t]
\centering
Kerr black hole excised outside of the horizon\par\medskip
\includegraphics[width=\textwidth]{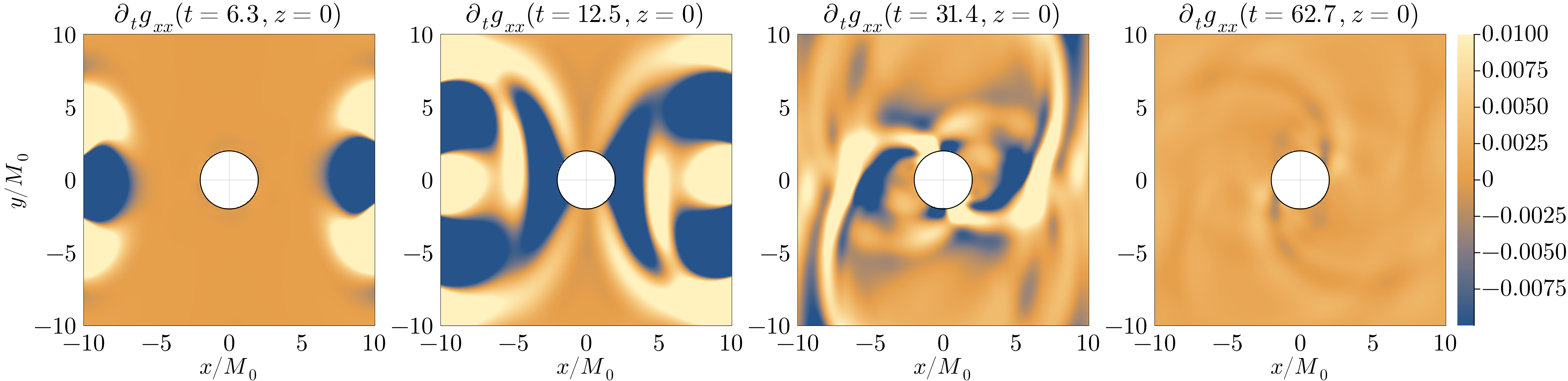}
Fractional Change in Quasi-Local Measures\par
\includegraphics[width=0.94\textwidth]{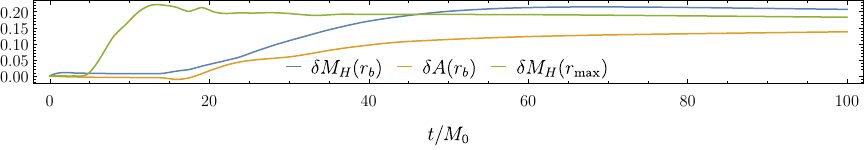}
\caption{Simulation of GW scattering on a Kerr black hole of initial mass $M_0=1$ and spin parameter $a=0.5M_0$. The black hole is excised \emph{outside} the horizon at $r_b=2M_0$, since the outer horizon is initially at $r_+\approx 1.87M_0$. Top panels show $\partial_t g_{xx}$ at several times on the $z=0$ slice. Bottom panel shows the fractional change in the surface area of the inner boundary $A(r_b)$ as well as the Hawking mass $M_H$ evaluated at $r_b$ and the largest sphere in the cubical domain $r_\mathrm{max}$. With the GW BCs we apply at the inner boundary (i.e. keeping these components constant), we see largely the same behavior from the quasi-local measures as the simulations excised inside the horizon, suggesting that these BCs lead to the absorption/transmission of GWs.}\label{Fig:Spin2}
\end{figure*}

\begin{figure}[h]
\includegraphics[width=\linewidth]{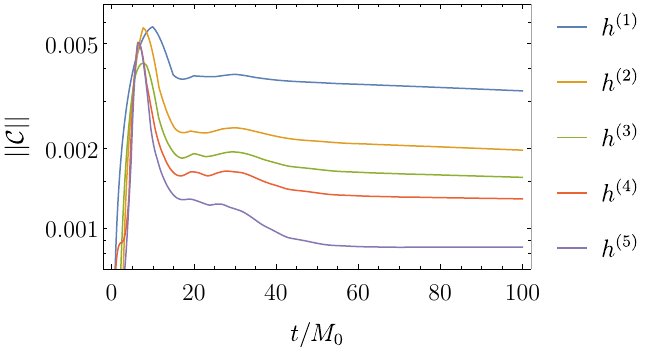}
\caption{Behavior of the gauge constraint norm over time for the five resolutions ($h^{(i)}$) in the Kerr black hole case with the boundary just outside the spinning black hole (${a=0.5M_0}$ and ${r_b=2M_0}$). We see that even though constraint violations are introduced when GWs are injected, they are damped out and settle to a finite value, which converges to zero with increased resolution. The other two simulations exhibit identical behavior in this respect.}\label{Fig:Constime}
\end{figure}

We track several quasi-local measures of mass and surface area (for numerical calculation details of closed surface integrals, see Appendix~\ref{App:Surfaces}). The surface area of the inner boundary is tracked because although the boundary is fixed in coordinate space, it may still physically grow/shrink. The black hole may also move into the computational domain if it is given linear momentum, but we inject GWs from two opposite sides at once to ensure this does not occur. The surface area of a coordinate sphere is defined as
\begin{align}
   A(r) = \oint_{S(r)} dS(r)\,,
\end{align}
where $S(r)$ is a coordinate sphere of radius $r$ and $dS(r)$ is the surface area element on that sphere. To measure the accretion of mass on the hole, we use the Hawking mass as a measure of the mass contained in a coordinate sphere
\begin{align}
   M_H(r) = \sqrt{\frac{A(r)}{16\pi}}\left(1+\frac{1}{16\pi}\oint_{S(r)}\rho\rho' dS(r)\right)\,,
\end{align}
where $\rho$ and $\rho'$ are the expansions of ingoing and outgoing null geodesics at the sphere, respectively. The integrand of the closed surface integral in this definition can also be expressed as \cite{Alcubierre:1138167}
\begin{align}
   \rho\rho'=(\sigma^{\mu\nu}K_{\mu\nu})^2-(\bar\nabla^i s_i)^2\,.
\end{align}
We also define the gauge constraint norm as
\begin{align}
    ||\sC||^2=\frac{\int m^{\mu\nu}\sC_\mu\sC_\nu dV}{\int dV}\,,
\end{align}
to track the convergence and damping of the constraint violating modes.

We present five different resolutions where ${h^{(i)} = 3h^{(i-1)}/4}$ as the resolution is increased by a factor of $4/3$ from one to the next starting at $h^{(1)}=0.25M_0$. The computational domain consists of a cube of volume $(20M_0)^3$ with an inner excised sphere of various radii, and we only consider grid spacings that are the same in all three coordinate directions.  We use a third order accurate Runge-Kutta algorithm to advance the equations with a time step $h_t=h/5$, which reflects the worst case CFL factor estimates from \cite{MATTSSON2017255}.




\subsection{Schwarzschild GW Scattering}

Here, we consider the initial condition of a Schwarzschild black hole in Kerr-Schild coordinates. This completely defines the metric $g^{\rm init}_{\mu\nu}$ on the initial slice and the gauge functions $H_\mu=\Gamma^{\rm init}_\mu$ we use for all time. Then we use our initial condition formalism to ensure $\sC^{\rm init}_\mu=\sC^{\rm init}_{i\mu\nu}=0$. The remaining components of $P_{\mu\nu}$ are set by requiring $\partial_t g^{\rm init}_{\mu\nu}=0$. We then inject GWs using the following model:
\begin{align}
   f_{\rm GW}(x^\alpha) = A_{GW}[r_e(x^\alpha)+1]^4[r_e(x^\alpha)-1]^4\,,\label{Eq:f-model}
\end{align}
defined in a region $r_e<1$ and is zero if $r_e\geq 1$ where 
\begin{align}
   r^2_e(x^\alpha)=[(x-x_0+v_x t)/r_x]^2+(y/r_y)^2+(z/r_z)^2\,.\label{Eq.Model}
\end{align}
This defines an ellipsoidal region with semi-axes $r_i$ that moves at $x$-directed velocity $v_x$ from its initial $x$-position (outside the computational domain) $x_0$. The amplitude $A_{GW}$ can then be tuned to add a desired amount of mass in GWs. We use $A_{GW}=130$, which results in an increase in Hawking mass of the hole by about 20\%. The polynomial in Eq.~(\ref{Eq:f-model}) ensures that the first three derivatives are continuous while allowing $f_{GW}$ to be exactly zero where $r_e\geq 1$. We inject this model in the following fashion
\begin{align}P\indices{_\mu_\nu^\alpha^\beta}U^{-\rm BC}_{\alpha\beta} = P\indices{_\mu_\nu^\alpha^\beta}U^{-\rm init}_{\mu\nu}(1+f_{GW})\,.
\end{align}
In this and following simulations, we use $r_y=r_z=10M_0$, $r_x=5M_0$, $v_x=\pm1$, and $x_0=(20\pm5)M_0$ for injections from the $\mp x$-directions. We then evolve the spacetime until the metric and quasi-local measures settle to approximately constant final values at $t=100M_0$. Results are depicted in Figure~(\ref{Fig:NoSpin}). We track the fractional change [$\delta f(t)\equiv(f(t)-f(0))/f(0)$] in both the Hawking mass of the inner sphere $M_H(r_b)$ as well as its value at the largest sphere in the domain $M_H(r_{\rm max})$. We also track the surface area of the inner sphere $A(r_b)$. As should be expected, $M_H(r_{\rm max})$ increases first, followed by $M_H(r_b)$, until they settle close to the same value at late time. The surface area $A(r_b)$ increases as well.

\subsection{Kerr GW Scattering}

In the same setup as the previous section, we use now a spinning black hole. We use a metric expressed in so-called Spherical Kerr-Schild coordinates defined in \cite{Chen_2021}, (see Appendix \ref{App:SKS}). This set of coordinates is transformed relative to traditional Kerr-Schild coordinates such that the horizons are spherical in coordinate space (as opposed to oblate ellipsoids). This allows us to place a spherical boundary at a constant coordinate distance from the horizon as we did in the previous section. We use spin $a=0.5M_0$ and the same inner boundary $r_b=1.5M_0$ which is still inside the outer horizon of the hole at $r_+\approx 1.87M_0$. We inject GWs in the same fashion as the Schwarzschild black hole and evolve the system until the metric and quasi-local measures settle to approximately constant final values at $t=100M_0$, with results depicted in Figure~(\ref{Fig:Spin}). This simulation exhibits largely the same behavior in quasi-local measures as the Schwarzschild case, except both $M_H(r_b)$ and $M_H(r_{\rm max})$ seem to increase slightly more by $t=100M_0$.

Next we showcase the same simulation, but instead with $r_b=2M_0$, just \emph{outside} of the outer horizon. The results are depicted in Figure~(\ref{Fig:Spin2}). When the hole is excised inside the horizon, $M_H(r_b)$ increases by 22\%, but when it is excised outside the horizon, $M_H(r_b)$ increases by 20\%, suggesting that the GW BC choice at the inner boundary, $h^{\rm BC}_{\mu\nu}=U^{-\rm init}_{\mu\nu}$, 
is a mostly transmitting/absorbing condition, with a reflection of about 10\% of the incident waves' mass (at least for this specific setup). The set of simulations we present are meant to showcase the stability we are able to achieve with the formalism and numerical methods we use when a boundary is placed in the strong gravity region of a dynamical spacetime, which has been notoriously difficult to achieve numerically in the past.

The convergence of the constraint norm for all three simulations is depicted in Figure~(\ref{Fig:Cons}), where the $D_{2-1}$ operator achieves less than second order convergence as is expected with SBP operators of this type. The behavior of the constraint norm over the period of the last simulation is depicted in Figure~(\ref{Fig:Constime}). The constraints increase as GWs are injected, but they soon damp out as expected, with increased resolution converging to zero at the final time.

\section{Future Work}

The numerical methods we demonstrate in this work allow for much more complex IBVPs that we plan to study further, including
\begin{itemize}
    \item Non-spherically shaped and non-convex inner boundaries
    \item Dynamical inner boundaries
    \item Non-cubical outer boundaries
    \item Higher accuracy operators
\end{itemize}
In particular, successful evolution of a dynamical boundary would allow for a much simplified binary black hole excision setup with finite differencing algorithms. This requires a way to initialize grid points that enter the computational domain as the boundary moves across the grid, which is usually done with extrapolation techniques. Once binary systems can be demonstrated, the addition of Cauchy characteristic matching would allow for the extraction of gravitational waveforms at infinity while preserving computational efficiency. Whether the boundary is placed inside or outside of an apparent horizon, its shape and movement over time should be dictated by an evolution formalism that ties the boundary to the black holes. We plan to study further how such an evolution can be achieved numerically when coupled with these embedded boundary methods.


We proposed in this work to use quasi-local conservation laws to define what it means to reflect gravitational radiation, as well as how to interpret what a given set of BCs implies about the mass, linear momentum, and angular momentum within a closed boundary as the system evolves over time. We plan to refine this concept into a proper mathematical framework that one could use to derive BCs for a desired boundary behavior. 

\section{Conclusion}

We have demonstrated successful black hole excision simulations both inside and outside of an apparent horizon on a strictly rectangular grid for the first time, which is the debut of embedded boundary finite differencing methods to the numerical relativity community. We have also shown that we can run stable numerical relativity simulations in a relatively small domain. This is important for Cauchy characteristic matching techniques, where the Cauchy evolution domain can in principle be made small, close to the scale of merging black holes, for example, to dramatically save on computational cost while retaining numerical stability and accurate gravitational waveforms at infinity. We presented a boundary condition framework based on controlling first derivatives of the metric in the generalized harmonic formulation, which has allowed us to cast Einstein's equations as an SBP-SAT scheme. This framework can be applied to other numerical methods common in the field, such as spectral methods, to obtain simulations that are numerically stable even when the computational domain has boundaries in the strong gravity regime. This marks a big leap toward our ultimate goal to evolve binary black hole simulations with boundary surfaces drawn just outside of the apparent horizons of the holes. This will allow us to impose quantum gravity motivated behavior on these boundary surfaces to obtain the first gravitational waveforms of black hole echoes from numerical relativity.

\begin{acknowledgments}

We are thankful for valuable discussions with Richard Epp, Luis Lehner, Sizheng Ma, Robert Mann, and other strong gravity researchers at the Perimeter Institute.
Numerical implementation for this project was accomplished with the Julia programming language \cite{Julia-2017} and the packages detailed in \cite{omlin2022highperformancexpustencilcomputations} and \cite{NakamuraTensorial2024}. This research was funded thanks in part to the Canada First Research Excellence Fund through the Arthur B. McDonald Canadian Astroparticle Physics Research Institute, the Natural Sciences and Engineering Research Council of Canada, and the Perimeter Institute. Research at Perimeter Institute is supported in part by the Government of Canada through the Department of Innovation, Science and Economic Development and by the Province of Ontario through the Ministry of Colleges and Universities.
\end{acknowledgments}

\appendix

\section{Discrete Closed Surface Integration}\label{App:Surfaces}

In order to calculate quasi-local quantities such as the Hawking mass, we require a way to approximate surface integrals on the embedded boundaries in our domain. This integration scheme must be compatible with the rectangular nature of the grid used in the numerical methods we consider in this work. One can express an integral over a convex closed surface $S$, defined by the function ${\phi(x^i)=0}$, with integrand $f(x^i)$, as two 2D region integrals:
\begin{align}
   \oint f dS &= \int_{X^\pm} f[x(y,z),y,z] \sqrt{\sigma}dydz\,,
\end{align}
where the region $X^\pm$ is the projection (or shadow) of the closed surface onto the $y$-$z$ plane and the $\pm$ indicates the inclusion of both the projections from the $\pm x$ sides of the surface. The function $x(y,z)$ is the expression $\phi(x^i)=0$ solved for $x$, which will have two solutions for a convex closed surface corresponding to the $\pm$ projections. The 2-metric determinant $\sigma$ is determined by projecting the 2-metric
\begin{align}
   \sigma_{ij}=\gamma_{ij}-s_i s_j\,,
\end{align}
onto the $y$-$z$ plane. The projection operator onto the 2-surface is given by
\begin{align}
   e^i_A=\frac{\partial x^i}{\partial\theta^A}\,.
\end{align}
In this example, we choose the  coordinates on the $y$-$z$ plane as $\theta^A=(y,z)$ so that ${e^i_A=(\partial_A x(y,z),\delta^y_A,\delta^z_A)}$. The projected 2D metric is given by
\begin{align}
   \sigma_{AB}=e^i_Ae^j_B\sigma_{ij}\,.
\end{align}
We then take the determinant to find the surface area element $\sqrt{\sigma}$ and integrate the integrand and the surface area element over the 2D regions $X^\pm$. An appropriate quadrature rule can then be chosen to perform this integral in approximation on a finite number of sampled points, as is defined for our numerical method in \cite{MATTSSON2017255}. Since we also have independent grid intersections with the $Y^\pm$ and $Z^\pm$ projection directions, we can repeat the surface integral three times, and average them to get a more accurate result.

\section{Spherical Kerr-Schild}\label{App:SKS}

A spinning black hole can be described in traditional Kerr-Schild coordinates, but this has the disadvantage that the horizons are oblate ellipsoids in coordinate space, which makes the problem of excision more complicated. However, with the aid of a coordinate transformation detailed in \cite{Chen_2021}, the horizons can instead take on a spherical shape in coordinate space. Using the coordinates $x^\mu=(t,x,y,z)$, this metric can be expressed in the usual Kerr-Schild form:
\begin{align}
   g_{\mu\nu} = \eta_{\mu\nu} + 2H l_\mu l_\nu\,,
\end{align}
with 
\begin{align}
   H = \frac{M r^3}{r^4+a^2 z^2}\,,\quad R^2=\frac{(r^4+ a^2 z^2)}{ (r^2+a^2)}\,,
\end{align}
and
\begin{align}
   l_\mu = \bigg(1,\frac{xR^2}{r^3}+\frac{ay}{r^2}\,\,,\frac{ yR^2}{r^3}-\frac{ax}{r^2}\,\,,\frac{ zR^2}{r^3}\bigg)\,.
\end{align}
While $\eta_{\mu\nu}$ is a representation of Minkowski space, it is not diagonal in these coordinates. It is however diagonal using a traditional transformation to spherical coordinates, so we can construct it with the following set of orthonormal basis vectors:
\begin{align}\
    t_\mu &=(1,\,0,\,0,\,0)\,,\\
    r_\mu &= (0,\,x,\,y,\,z)/A\,,\\
    \theta_\mu &= (0,\,-xz,\,-yz,\,x^2+y^2)/B\,,\\
    \varphi_\mu &= (0,\,-y,\,x,\,0)/C\,,
\end{align}
with
\begin{align}
    A^2 = \frac{r^4(r^2+a^2)}{r^4+a^2z^2}\,,
    B^2 = \frac{r^6(x^2+y^2)}{r^4+a^2z^2}\,,
    C^2 = \frac{r^2(x^2+y^2)}{r^2+a^2}\,,
\end{align}
and
\begin{align}
    \eta_{\mu\nu}=-t_{\mu}t_{\nu}+r_\mu r_\nu+\theta_\mu\theta_\nu+\varphi_\mu\varphi_\nu\,,
\end{align}
taking care that a smooth limit to the diagonal metric ${\eta_{\mu\nu}=\mathrm{diag}(-1,1+(a/z)^2,1+(a/z)^2,1)}$ is attained when $x=y=0$.

\bibliography{refs}

\begin{thebibliography}{10}

\bibitem{Dailey_2023}
Conner Dailey, Niayesh Afshordi, and Erik Schnetter.
\newblock Reflecting boundary conditions in numerical relativity as a model for black hole echoes.
\newblock {\em Classical and Quantum Gravity}, 40(19):195007, aug 2023.

\bibitem{Oshita:2019sat}
Naritaka Oshita, Qingwen Wang, and Niayesh Afshordi.
\newblock On reflectivity of quantum black hole horizons.
\newblock {\em Journal of Cosmology and Astroparticle Physics}, 2020(04):016--016, apr 2020.

\bibitem{Wang:2018gin}
Qingwen Wang and Niayesh Afshordi.
\newblock {Black hole echology: The observer\textquoteright{}s manual}.
\newblock {\em Phys. Rev. D}, 97(12):124044, 2018.

\bibitem{Ikeda:2021uvc}
Taishi Ikeda, Massimo Bianchi, Dario Consoli, Alfredo Grillo, Jos\`e~Francisco Morales, Paolo Pani, and Guilherme Raposo.
\newblock {Black-hole microstate spectroscopy: Ringdown, quasinormal modes, and echoes}.
\newblock {\em Phys. Rev. D}, 104(6):066021, 2021.

\bibitem{Wang:2019rcf}
Qingwen Wang, Naritaka Oshita, and Niayesh Afshordi.
\newblock {Echoes from Quantum Black Holes}.
\newblock {\em Phys. Rev. D}, 101(2):024031, 2020.

\bibitem{Abedi:2018npz}
Jahed Abedi and Niayesh Afshordi.
\newblock Echoes from the abyss: a highly spinning black hole remnant for the binary neutron star merger {GW}170817.
\newblock {\em Journal of Cosmology and Astroparticle Physics}, 2019(11):010--010, nov 2019.

\bibitem{Abedi:2020sgg}
Jahed Abedi and Niayesh Afshordi.
\newblock Echoes from the abyss: A status update, 2020.

\bibitem{Abedi:2016hgu}
Jahed Abedi, Hannah Dykaar, and Niayesh Afshordi.
\newblock {Echoes from the Abyss: Tentative evidence for Planck-scale structure at black hole horizons}.
\newblock {\em Phys. Rev. D}, 96(8):082004, 2017.

\bibitem{Burgess:2018pmm}
C.~P. Burgess, Ryan Plestid, and Markus Rummel.
\newblock Effective field theory of black hole echoes.
\newblock {\em Journal of High Energy Physics}, 2018(9), sep 2018.

\bibitem{Cardoso:2019apo}
Vitor Cardoso, Valentino~F. Foit, and Matthew Kleban.
\newblock Gravitational wave echoes from black hole area quantization.
\newblock {\em Journal of Cosmology and Astroparticle Physics}, 2019(08):006--006, aug 2019.

\bibitem{Ma_2024}
Sizheng Ma, Jordan Moxon, Mark~A. Scheel, Kyle~C. Nelli, Nils Deppe, Marceline~S. Bonilla, Lawrence~E. Kidder, Prayush Kumar, Geoffrey Lovelace, William Throwe, and Nils~L. Vu.
\newblock Fully relativistic three-dimensional cauchy-characteristic matching for physical degrees of freedom.
\newblock {\em Physical Review D}, 109(12), June 2024.

\bibitem{McGrath_2012}
Paul~L McGrath, Richard~J Epp, and Robert~B Mann.
\newblock Quasilocal conservation laws: why we need them.
\newblock {\em Classical and Quantum Gravity}, 29(21):215012, October 2012.

\bibitem{EC_Formulation}
Lawrence Kidder, Mark Scheel, Saul Teukolsky, Eric Carlson, and Gregory Cook.
\newblock Black hole evolution by spectral methods.
\newblock {\em Physical Review D}, 62(8), sep 2000.

\bibitem{Garfinkle2002}
David Garfinkle.
\newblock Harmonic coordinate method for simulating generic singularities.
\newblock {\em Phys. Rev. D}, 65:044029, Jan 2002.

\bibitem{Szil_gyi_2003}
Béla Szilágyi and Jeffrey Winicour.
\newblock Well-posed initial-boundary evolution in general relativity.
\newblock {\em Physical Review D}, 68(4), August 2003.

\bibitem{Friedrich1985}
Helmut {Friedrich}.
\newblock {On the hyperbolicity of Einstein's and other gauge field equations}.
\newblock {\em Communications in Mathematical Physics}, 100(4):525--543, December 1985.

\bibitem{Lindblom_2006}
Lee Lindblom, Mark~A Scheel, Lawrence~E Kidder, Robert Owen, and Oliver Rinne.
\newblock A new generalized harmonic evolution system.
\newblock {\em Classical and Quantum Gravity}, 23(16):S447, jul 2006.

\bibitem{Brown_2011}
J.~David Brown.
\newblock Action principle for the generalized harmonic formulation of general relativity.
\newblock {\em Physical Review D}, 84(8), October 2011.

\bibitem{SBP_Review}
David~C. {Del Rey Fernández}, Jason~E. Hicken, and David~W. Zingg.
\newblock Review of summation-by-parts operators with simultaneous approximation terms for the numerical solution of partial differential equations.
\newblock {\em Computers \& Fluids}, 95:171--196, 2014.

\bibitem{Shifted_Wave}
Ken Mattsson and Florencia Parisi.
\newblock Stable and accurate second-order formulation of the shifted wave equation.
\newblock {\em Communications in Computational Physics}, 7, 01 2009.

\bibitem{MATTSSON2017255}
Ken Mattsson and Martin Almquist.
\newblock A high-order accurate embedded boundary method for first order hyperbolic equations.
\newblock {\em Journal of Computational Physics}, 334:255--279, 2017.

\bibitem{Mishra2010}
Siddhartha Mishra and Magnus Svärd.
\newblock On stability of numerical schemes via frozen coefficients and the magnetic induction equations.
\newblock {\em BIT}, 50:85--108, 03 2010.

\bibitem{Gundlach_2006}
Carsten Gundlach and José~M Martín-García.
\newblock Hyperbolicity of second order in space systems of evolution equations.
\newblock {\em Classical and Quantum Gravity}, 23(16):S387–S404, July 2006.

\bibitem{Pretorius_2005}
Frans Pretorius.
\newblock Evolution of binary black-hole spacetimes.
\newblock {\em Physical Review Letters}, 95(12), September 2005.

\bibitem{Pretorius_2005_2}
Frans Pretorius.
\newblock Numerical relativity using a generalized harmonic decomposition.
\newblock {\em Classical and Quantum Gravity}, 22(2):425–451, January 2005.

\bibitem{PhysRevD.107.064013}
Jordan Moxon, Mark~A. Scheel, Saul~A. Teukolsky, Nils Deppe, Nils Vu, Francois H\'ebert, Lawrence~E. Kidder, and William Throwe.
\newblock Spectre cauchy-characteristic evolution system for rapid, precise waveform extraction.
\newblock {\em Phys. Rev. D}, 107:064013, Mar 2023.

\bibitem{Alcubierre:1138167}
Miguel Alcubierre.
\newblock {\em {Introduction to 3+1 numerical relativity}}.
\newblock International series of monographs on physics. Oxford Univ. Press, Oxford, 2008.

\bibitem{PhysRevD.71.064020}
Lawrence~E. Kidder, Lee Lindblom, Mark~A. Scheel, Luisa~T. Buchman, and Harald~P. Pfeiffer.
\newblock Boundary conditions for the einstein evolution system.
\newblock {\em Phys. Rev. D}, 71:064020, Mar 2005.

\bibitem{Babiuc_2007}
M.~C. Babiuc, H-O. Kreiss, and Jeffrey Winicour.
\newblock Constraint-preserving sommerfeld conditions for the harmonic einstein equations.
\newblock {\em Physical Review D}, 75(4), February 2007.

\bibitem{Rinne_2007}
Oliver Rinne, Lee Lindblom, and Mark~A Scheel.
\newblock Testing outer boundary treatments for the einstein equations.
\newblock {\em Classical and Quantum Gravity}, 24(16):4053–4078, July 2007.

\bibitem{SHARAN2022111341}
Nek Sharan, Peter~T. Brady, and Daniel Livescu.
\newblock High-order dimensionally-split cartesian embedded boundary method for non-dissipative schemes.
\newblock {\em Journal of Computational Physics}, 464:111341, 2022.

\bibitem{Chen_2021}
Yitian Chen, Nils Deppe, Lawrence~E. Kidder, and Saul~A. Teukolsky.
\newblock Efficient simulations of high-spin black holes with a new gauge.
\newblock {\em Physical Review D}, 104(8), October 2021.

\bibitem{Julia-2017}
Jeff Bezanson, Alan Edelman, Stefan Karpinski, and Viral~B Shah.
\newblock Julia: A fresh approach to numerical computing.
\newblock {\em SIAM {R}eview}, 59(1):65--98, 2017.

\bibitem{omlin2022highperformancexpustencilcomputations}
Samuel Omlin and Ludovic Räss.
\newblock High-performance xpu stencil computations in julia, 2022.

\bibitem{NakamuraTensorial2024}
Keita Nakamura.
\newblock Tensorial.jl: a {J}ulia package for tensor operations, 2024.

\end{thebibliography}

\end{document}